\documentclass[aps,prc,twocolumn,showpacs,amsmath,amssymb]{revtex4}

\usepackage{graphicx}

\begin{document}


\title{Pentaquark as Kaon--Nucleon Resonance}

\author{D.~E.~Kahana}
\author{S.~H.~Kahana}

\affiliation{Physics Department, Brookhaven National Laboratory,
   Upton, NY 11973, USA}

\date{\today}  
 
\begin{abstract}

Several recent  experiments have  reported evidence for  a narrow
feature in  the $K^+$-neutron system, an  apparent resonant state
$\sim 100$  MeV above threshold and  with a width $  \le 25$ MeV.
This state has been labelled as $\Theta^+$ (previously as $Z^*$),
and because of the implied  inclusion of a anti-strange quark, is
referred to as a pentaquark, that is, five quarks within a single
bag.  We present an alternative explanation for such a structure,
as a higher  angular momentum resonance in the  isospin zero $K^+
N$ system.   One might call this  an exit channel  or a molecular
resonance.   In  a non-relativistic  potential  model  we find  a
possible  candidate  for the  kaon-nucleon  system with  relative
angular  momentum  $L=3$,  while  $L=1$ and  $2$  states  possess
centrifugal barriers too low to confine the kaon and nucleon in a
narrow  state at  an energy  so high  above threshold.   A rather
strong state-dependence  in the potential  is essential, however,
for eliminating an observable $L=2$ resonance at lower energies.

\end{abstract}


\maketitle 

Several  groups ~\cite{spring,diana,jlab,elsa} have  reported the
presence of a resonance, referred to presently as the $\Theta^+$,
in a  variety of experimental  configurations.  This experimental
situation  stands  in sharp  contrast  to  that  relative to  the
6-quark  H-dibaryon~\cite{jaffe1}, which  at first  glance seemed
theoretically  to be a  candidate for  an exotic  state eminently
more likely to  be discovered. For the $H$  there have been many,
so far fruitless searches.   We offer elaborations on some points
made  by Capstick, {\it  et~al.}~\cite{capstick}, on  the general
nature  of   resonances  in  the  kaon-nucleon   system  and  the
connection  between their  angular momenta,  widths  and energies
above threshhold.

Perhaps the simplest data comes  from the SAPHIR detector at ELSA
~\cite{elsa},  wherein  the  positive strangeness  $\Theta^+$  is
photo-produced  off  a simple  proton  target.   The final  state
contains $nK^+K_s^0$ and the relevant system is identified in the
missing   mass  spectrum   of  the   $K_s^0$.   The   $K_s^0$  is
reconstructed  from its  two $\pi$  decay, preferentially  in the
forward direction.   These authors conclude the  $\Theta^+$ is an
isoscalar due to the absence  of a $\Theta^{++}$ in the $\gamma p
\rightarrow    pK^+K^-$   channel.    Interestingly,    in   this
preferentially forward detection  of the $K_s^0$~\cite{elsa}, the
transfer  of   appreciable  angular  momentum   to  the  observed
resonance would be favored.

Previously       and      presently,       several      theorists
~\cite{praszalowicz,walliser,diakonov,weigel,karliner}       have
predicted  the  existence  of   such  states,  on  occasion  with
astonishing  accuracy~\cite{diakonov} one  might  add, using  the
Skyrme  model  topological  soliton  ~\cite{skyrme} as  a  basis;
others,       including       the       experimental       groups
themselves,~\cite{spring,diana,jlab,jaffe,capstick}          have
described  the  state  as  a  bag~\cite{mitbag}  containing  five
quarks, {\it viz.}  a pentaquark.
 
There are  perhaps some difficulties inherent in  either of these
theoretical approaches, since  a $^1P$ configuration is generally
selected  as the  primary candidate.   There is  a wealth  of low
energy          $K           N$          scattering          data
~\cite{sakitt,dover,martin,bgrt,roper}   which  presumably  would
have revealed a low-angular  momentum $\Theta^+$ resonance in the
total cross-section or elsewhere.   This problem could be avoided
if the state possessed a  truly narrow width, perhaps $\Gamma \le
3$ MeV  ~\cite{newanal}.  On the  other hand if a  higher angular
momentum   $L=2,3$  resonance   were  sought,   the   absence  of
observations in  $\sim 100$ MeV scattering data  might perhaps be
bypassed ~\cite{newanal}.

Pentaquark  bag  and  soliton   solutions  both  face  a  related
problem. They will,  in decay, be connected to  the outgoing tate
via   a   low  angular   momentum,   molecular  `doorway'   state
~\cite{capstick}. At $100$ MeV  of excitation, an $L=1$ resonance
could  acquire so  large a  width as  to be  non-existent  in the
absence of $SU_f(3)$ symmetry  breaking, or, in nuclear structure
terms the presence of  a small `spectroscopic' factor to suppress
the decay ~\cite{capstick,nussinov}.
  
In searching for a suitable phenomenological `effective' $K$--$N$
potential one  finds that a  resonance of width $7$--$21$  MeV at
the rather elevated excitation energy  of $100$ MeV can indeed be
constructed. Such an  effective potential presumably represents a
final state interaction between  the kaon and nucleon components,
{i.~e.}~a  molecule  like  doorway  state.   Proceeding  in  this
fashion  may  amount to  trading  one  problem  for another,  and
involves an appreciable degree  of artificiality. But in doing so
we can throw some light on the experimental problems arising from
the  assumption  that  the  $\Theta^+$ possesses  a  low  angular
momentum.

It is in fact already  known, generally from phase shift analyses
of data ~\cite{dover,martin,bgrt}, that the $KN$ system possesses
a  considerable  state dependence  in  its  various  $S$ and  $P$
configurations.  In particular the isoscalar singlet $P$ seems to
evidence the greatest attraction  and in some modeling, the phase
shift passes upwards through $\pi/2$, however, at the rather high
relative momentum of $800$  MeV/c ~\cite{bgrt,bnl}.  In this very
brief  contribution  we only  attempt  to  delineate the  minimum
requirements  for a molecular  resonance to  exist, $K^+$  + $N$,
rather than a single bag.

A possible candidate state with an appropriate and natural single
particle width-energy  relation will, as we  shall see, possesses
$L=3$. If  there is no  dependence of the effective  potential on
orbital angular momentum, it is likely the $L=2$ channel also has
a resonant state, which will  generally be narrower and closer to
threshold.  Thus  an $L=2$ resonance remains  a possibility, with
perhaps  a  reduced   cross-section  for  production  considering
somewhat large momentum transfer required.

In this  picture, we have in  mind, first the creation  of a very
short-lived seven-quark 'bag' and soon after the fission into the
final $\Theta^+$  and $K_s^0$.  Both  the initial object  and the
$\Theta^+$  may  well  be  deformed~\cite{mitbag,iachello}.   The
$\Theta^+$-molecule  is  likely  prolate  before  its  subsequent
fission  (decay) into its  $K$ and  $N$ components.   We envision
that the interaction between  these components is mainly on their
surface, perhaps a particle-vibrational coupling.
  
Explicit calculations  were done using  the non-relativistic code
GAMOW due to T.~Verse, K.~F.~Pal and Z.~Balogh ~\cite{resonance},
in  its  original  form  and   also  in  a  version  modified  to
essentially  exclude interactions  inside  some radius,  somewhat
inside the $0.8$--$1$ fm expected  for the size of a nucleon bag.
Non-relativistic kinematics should be adequate for $K$-$N$ system
at the  rather low relative energy  of $100$ MeV  ascribed to the
$\Theta^+$.  From the outset, it is abundantly clear that a state
of  such energy  and  width  cannot be  easily  sustained by  the
centrifugal barrier in  states with $L \le 2$.   This could be an
advantageous  circumstance,   should  further  experiments  truly
eliminate the presence of even  a narrow $\Theta^+$ in low energy
$KN$  scattering.   An  $L=3$  resonance  would  have  been  very
difficult to see in existing $100$ MeV data ~\cite{newanal}.  The
widths  considered here are  essentially what  is referred  to as
single-particle in  nature and  ignore for now  narrowing arising
from any symmetry-breaking or other mechanism.

One  potential  which can  produces  a  resonance  in the  higher
partial waves is a surface Saxon-Woods form:

\begin{equation}
V_{surface}(r)= -V^s_0\left[\frac{4e^{(r-R)/a}}{(1+e^{(r-R)/a})
^2}\right].
\end{equation}
The selection  of a surface interaction, sharply  peaked at that,
is  discussed  below.   If   one  wants  to  limit  the  produced
resonances  to say  $L=2,3,4$, some  appreciable degree  of state
dependence must  be introduced.  One  would also want to  cut off
this  dependence in  even  higher partial  waves.  The  potential
stength then appears as,
\begin{equation}
V^s_0= A[\alpha +\beta F(L)],
\end{equation}

There  is no  problem  introducing {\it  some} state  dependence,
certainly     for     isospin;     most    earlier     treatments
~\cite{dover,martin}  present  the   $S$  and  $P$  states  quite
differently.     More   interestingly,   Hasenfratz    and   Kuti
~\cite{mitbag,hasenfratz,tomio,bohr}   suggested   that  isolated
quark bags  may be  treated like liquid  drops, in  close analogy
with  nuclei.   In  such   a  dynamical  treatment,  the  bag  is
deformable, and susceptible to  surface oscillations which can be
expanded in spherical harmonics,  each characterized by a quantum
number   $l$.     One   can   imagine,   as    one   example,   a
particle-vibrational coupling producing the effective interaction
in Eq.(1), by a quark in  say the kaon coupling to a vibration of
the  nucleon  surface, thus  our  choice  of  $R$.  These  latter
authors~\cite{hasenfratz}  point out  that the  surface potential
energy is proportional to

\begin{equation}
c_l= (l-1)(l+2)\rho_0^2\sigma,
\end{equation}
where  $\rho_0^2$ is  the  bag radius  and  $\sigma$ the  surface
tension.   Such  a  coupling  might  then lead  to  a  comparable
dependence  in  the  overall  effective $K$-$N$  interaction  via
particle-vibrational  coupling.   The  number  of  surface  waves
contributing  is naturally  cut  off  at higher  $l$  by the  the
underlying microscopic structure of  the bag surface, thus only a
very few surface  waves need enter. From Eq(3)  the $l=1$ mode is
absent and it  is not unreasonable, given that  only five valence
quarks are present~\cite{bohr}, that we assume only relative D, F
and G  waves are affected.  Even  an $l=4$ surface  wave may have
too  small  a wave  length  for  consideration  given that  three
valence quarks are present in the nucleon.

The  particle-vibrational  coupling in  nuclei  is,  in any  case
certainly  strongly  surface  peaked,  with  a  form  factor$\sim
rdV^{vol}/dr$ ~\cite{bohr}.  One then expects the diffusivity $a$
to  be quite  small. We  might  assume for  the surface  strength
a form like:

\begin{equation}
V^S_0= A[\alpha + \beta(L-1)(L+n)],
\end{equation}
which is, of course, not  directly justified and should be viewed
as  a phenomenological ansatz.   Eq(4) implies  no effect  of the
particle-vibrational coupling  in the $L=1$ system  and adds some
further  orbital  dependence.   Since  we  are  considering  only
$L=2,3,4$  states  here,  we   could  simply  quote  results  for
differing ratios of potential strengths in these orbits, ignoring
in effect the explicit choices made in this equation.

One can argue further that the simple kaon-nucleon picture should
be  modified  to account  for  a  bag-like  interior, say  within
separations $\sim 1.0$ fm,  with the molecular state arising from
interactions  at larger  distance.  So  a better  model  would be
hybrid in nature, its states  being mixtures of an interior quark
bag and  exterior hadrons.  The surface  potential employed here,
with  a small  diffusivity,  achieves this  to  some extent,  but
perhaps  the interior  structure is  more complicated.   A rather
obvious, but nevertheless arbitrary, change would be to introduce
a constant volume potential fixed to cancel the centrifugal force
just  inside the assumed  surface and  thus achieve  some average
cancellation in the interior,  or alternatively to just eliminate
this latter interaction inside.   The first procedure reduces the
required  required strength of  the surface  interaction $V^S_0$;
the   second   narrows    the   relevant   $L$-state   resonances
considerably.

The radius parameter in  Eq(1) is taken as $R=1.1(1.0087)^{1/3}fm
=1.130$ fm and the diffusivity  as $a=0.2$ fm.  The narrowness of
the surface  well, {i.~e.}~the small  value of $a$,  implies that
the  interaction  is not  strong  inside  or  outside the  chosen
surface separation.  The choice of diffusivity greatly influences
the resonance width; $a$ and $\Gamma$ both decrease together.

Specifically,  using  a potential  with  no interior  centrifugal
force and setting $\alpha=0$, $n=0$, $\beta=1$ and $A=70.835$ MeV
yields an $L=3$, $KN$  state at $\epsilon=103-3.76i$ MeV.  For an
unaltered interior potential with  $A=134.2$ MeV, the $L=3$ state
appears at  $\epsilon=100 - 11.2i$ MeV.  The  widths $\Gamma$ are
twice  the imaginary  part of  the  energies.  In  all cases  the
radial  integral of  the  surface potential,  absent the  angular
factor $4\pi$, is near to or less than $50$ Mev-fm$^3$.  Moreover
this  applies   to  only  a  few  partial   waves.   The  usually
anticipated  weakness of the  $K^+$--$N$ interaction~\cite{dover}
to some extent justifies setting $\alpha=0$, or small, in Eq(4).

For  $\alpha=n=0$ in  Eq(4),  the $L=2$  strength  is reduced  by
$1/3$.  With any of the  models, even for less drastic reductions
in  the $L=2$  strength, this  implies either  that  no $D$-state
resonance exists, or  that the state is present  at some $84$ MeV
(no interior centrifugal force) but with a width $\Gamma \sim 80$
MeV.  On  the other  hand, if the  surface potential  strength is
kept  at  the value  for  $L=3$  of  $6(70.83)=425$ MeV,  then  a
resonance develops in $L=2$  at $\epsilon=35.4-4.5i$ MeV.  Such a
$D$ state,  closer to threshold,  is perhaps lower  in production
cross-section and more difficult to detect.  Here we keep in mind
the somewhat large momentum  tranfer occasioned by the production
of at least one $K$ meson in the final state.

In all situations the above  $L=3$ surface potential alone is too
weak to generate an $L=4$  resonance near $100$ MeV. We expect in
any case some  cutoff effect to start with  higher partial waves,
probably by $L=4$, since the nucleon and kaon possess few valence
quarks.    Similar  results  could   have  been   obtained  using
energy-dependent potentials, but as  we pointed out there is some
justification for a particle-vibrational coupling acting in a few
surface states.

This  exercise  certainly  has  artificial aspects,  notably  the
state-dependent  potentials, although  such dependences  are just
what one  would expect to arise  in coupling the  meson quarks to
the nucleon  bag surface.  At  first sight the  surface potential
depths appear large, however, since the diffusivity is small, the
integrated  moment   of  the   potential,  which  has   a  direct
significance  in producing states  is in  fact quite  small.  One
might well  ask why such  dynamics are limited to  $K^+$--$N$ and
not also  present in $K^-$--$N$.   Two answers are  possible: the
latter  state is  not exotic,  proceeds through  $K^-$ absorption
involving known and observed $Y^*$ resonances in the $s$-channel,
also,  the $K^+N$ system  generally involves  weaker interactions
and may  then more easily  exhibit the surface  effects discussed
above.

The approach  also has its advantages:  the apparent experimental
absence of the coupling of  an assumed low angular momentum state
to low  energy $K$--$N$ channels,  and the natural  appearance at
$\sim  100$ MeV of  an $L=3$  state with  about the  right width.
Should  closer  experimental  study  reveal  another,  say  $L=2$
resonance  closer  to  threshold,  one  would have  to  take  our
approach more  seriously.  If the actual width  of the $\Theta^+$
proved to  be truly small $\le  1$ MeV, then  an explanation must
probably be sought in some other model.

Finally one should note  the rather large cross-section, found by
the  SAFIR   collaboration  ~\cite{elsa},  $\sim   300$  nb,  for
production of the $\Theta^+$. As we noted above, the large change
in mass, {i.~e}~the  production of one or two  $K$'s in the final
state,  and the  forward detection  of the  $K_s^0$,  likely also
favor an appreciable transfer of angular momentum to the putative
$\Theta$ resonance.   This is especially clear if  in $\gamma p$,
the  production of  the final  state proceeds  through  a doorway
$N^*$.   The  nucleon form  factor  then  enters  the $\gamma  p$
cross-section  roughly  as  $|f_N[R_p  2M(K)]|^2$, while  if  the
$K_s^0$ is produced  at the first $\gamma p$  vertex then perhaps
only a  single unit of M(K)  is present.  Such  effects are seen,
for  example,   in  $(\pi^+,K^+)$  reactions   on  nuclei,  which
preferentially produce high  angular momentum hypernuclear levels
~\cite{pi+k+}.

To repeat: one  lesson to be learned from  this examination of an
effective  potential  model,  concerns the  relationship  between
resonance energy  and single particle  width. It is  difficult to
obtain a narrow $P$-state  at $100$ Mev without symmetry breaking
or other weakening of the decay.  Capstick {et.~al.}~have already
pointed this out and one can point to the strong but broad $L=1$,
$\Delta$ as a further example of it.  The centrifugal force $\sim
1/\mu$ sets  the scale in any given  system, $\pi$--$N$ naturally
having   a   higher  energy   scale   than  $K$--$N$.    Nussinov
{et~al.}~\cite{nussinov}  discuss alternate  means  for narrowing
the $\Theta^+$  and related resonances.   Should the as  yet only
weakly constrained  width of the  $\Theta^+$, at $\Gamma  \le 25$
MeV,   actually   prove  to   be   considerably  narrower   Refs.
~\cite{capstick,nussinov} become even more relevant.

Capstick  {et~al.}~have  cleverly   narrowed  the  width  of  the
$\Theta^+$ by  taking this object  to be an isotensor.   They are
then  faced with  the  existence  of the  other  members of  this
multiplet,  in particular the  $\Theta^{++}$.  Our  suggestion is
clearly  highly   phenomenological,  and  a   purely  final-state
kaon-nucleon potential  model cannot  be the whole  answer, since
very short  distances must be  described by the  underlying quark
nature  of  hadrons.   Nevertheless,  the relevance  of  somewhat
higher orbital states cannot be eliminated immediately.  Clearly,
further experimental study of these exotic objects, in particular
of their angular  momenta, and of the entire  low energy $K$--$N$
system, is both interesting and necessary.

The  authors  are  grateful  to D.~J.~Millener  for  many  useful
discussions.

This  manuscript  has  been  authored  under  the  US  DOE  grant
NO. DE-AC02-98CH10886.


\begin{thebibliography}{99}

\bibitem{spring}

T.~Nakano  \textit{et al}, [LEPS  Collaboration], Phys.~Rev.~Lett
\textbf{91} 012002 (2003).

\bibitem{diana}

V.~V.~Barmin     \textit{et    al},     [DIANA    Collaboration],
arXiv:hep-ex/0304040.


\bibitem{jlab}

S.~Stepanyan \textit{et al} [CLAS  Collaboration], arXiv:
hep-ex/0303018.


\bibitem{elsa}

J.~Barth \textit{et al}, [SAPHIR Collaboration]
arXiv:hep-ex/0307083.

\bibitem{jaffe1}
R.~L.~Jaffe, Phys.~Rev.~Lett \textbf{38},195 (1977).


\bibitem{capstick}
S.~Capstick, P.~R.~Page and W.~Roberts, arXiv:hep-ph/0307019.


\bibitem{praszalowicz} 
M.~Praszalowicz, Proceedings, World Scientific, Singapore 1987,
M.~Jezabek, M.~Praszalowicz, (Eds.),  111; arXiv:hep-ph/0308114


\bibitem{walliser}

H.~Walliser,    Nucl.~Phys.   \textbf{A548},    649   (1992);
H.~Walliser and V.~Kopeliovitch, arXiv:hep-ex/0307019.



\bibitem{diakonov}

D.~Diakonov, V.~Petrov, and M.~Polyakov, Z.~Phys. \textbf{A359}, 
305 (1970).


\bibitem{weigel} 
H.~Weigel, Eur.~Phys.~J. \textbf{A2} 391 (1998).

\bibitem{karliner}
M.~Karliner and H.~J.~Lipkin, arXiv:hep-ph/0307243.


\bibitem{skyrme}

T.~H.~Skyrme,     Nucl.~Phys. \textbf{31}, 556 (1962);
Proc.~Roy.~Soc. Lond. A \textbf{260},127 (1961).

\bibitem{jaffe}
R.~L.~Jaffe, F.~Wilczek, arXiv:hep-ph//0307341.

\bibitem{mitbag}
A.~Chodos, R.~L.~Jaffe, K.~Johnson, C.~B.~Thorn and V.
~F.~Weisskopf, Phys.~Rev. D \textbf{9}, 3471 (1974).

\bibitem{sakitt}

M.Sakitt, J.~Skelly and J.~A.~Thompson, Phys.~Rev. D 
\textbf{12}, 3386 (1975); Phys.~Rev. D \textbf{15}, 
1846 (1977).

\bibitem{dover}

C.~B.~Dover and G.~E.~Walker, Phys.~Rep \textbf{89} No.1, 
(1982).

\bibitem{martin}
B.~R.~Martin, Nucl.~Phys. B \textbf{94} 413 (1975).

\bibitem{bgrt}

G.~Giacomelli \textit{et al} [BGRT Collaboration], Nucl.~Phys. B 
\textbf{71} 138 (1974).

\bibitem{bnl}
A.~S.~Carroll \textit{et al}, Phys.~Lett.  B \textbf{45}, 531 
(1973); R.~J.~Abrams \textit{et al}, Proc.~Int.Conf. on High Energy 
Physics, Batavia, 1972.

\bibitem{roper}

J.~S.~Hyslop, R.~A.~Arndt, L.~D.~Roper, and R.~L.~Workman, 
Phys.~Rev.~D, \textbf{46}, 961 (1992).


\bibitem{newanal}

R.~A.~Arndt, I.~J.~Strakovsky and R.~L.~Workman, arXiv:
nucl-th/0308012.

\bibitem{nussinov}

S.~Nussinov, arXiv:hep-ph/0307357; R.~Gothe and S.~Nussinov, 
arXiv:hep-ph/0308230.

\bibitem{iachello}           R.~Bijker,~F.~Iachello,~A.~Leviathan,
Annal.~Phys. \textbf{236}, 69 (1994).

\bibitem{resonance}

T.~Vertse,    K.~F.~Pal     and    Z.~Balogh,    Comp.~    Phys.~
Comm. \textbf{27}, 309 (1982).

\bibitem{hasenfratz}
P.~Hasenfratz and J.~Kuti, Phys.~Rep. \textbf{40}, 83 (1978);

\bibitem{tomio}
L.~Tomio and Y.~Nogami, Phys.~Rev. D \textbf{31}, 2818 (1985).

\bibitem{bohr}
A.~Bohr and B.~Mottelson, in  ``Nuclear Structure'', Vol.2 
[W.~A.~Benjamin, INC, Reading Massachusetts, 1975].


\bibitem{pi+k+}
D.~J.~Millener, A.~I.~P. Conference Proceedings \textbf{224},
Particles and Fields, Series \textbf{43}, 129 (1990); R.~E.
~Chrien \textit{et al}, Nucl.~Phys. A \textbf{478}, 705c 
(1988).

   
\end{thebibliography}
\end{document}